\documentclass[letterpaper,12pt]{article}
\pdfoutput=1

\usepackage{jheppub}
\usepackage{hyperref}
\usepackage{amsmath}
\usepackage{amsfonts}
\usepackage{graphicx}
\usepackage{caption}
\usepackage{subcaption}
\usepackage{tikz}
\usepackage[english]{babel}

\usepackage{array}
\usepackage{mathtools}
\def\beq{\begin{equation}}
\def\eeq{\end{equation}}
\def\bal{\begin{align}}
\def\eal{\end{align}}

\newcommand{\Tr}{\ensuremath{\mathrm{Tr}\,}}

\newcommand{\MCG}{\ensuremath{\mathrm{MCG}}}

\newcommand{\Lint}{\ensuremath{\mathcal{L}_\mathrm{int}}}

\title{
The Cut Equation
} 

\author[a]{N.~Arkani-Hamed,}\author[a]{H.~Frost,}\author[a,b]{G.~Salvatori}

\affiliation[a]{School of Natural Sciences, Institute for Advanced Study, Princeton, NJ, 08540, USA}
\affiliation[b]{Max-Plank-Instit\"ut fur Physik, Werner-Heisenberg-Institut, D–85748 Garching bei M\"unchen, Germany}
\emailAdd{arkani@ias.edu}
\emailAdd{salvatori@ias.edu}
\emailAdd{frost@ias.edu}

\date{\today}

\abstract{Scattering amplitudes for colored theories have recently been formulated in a new way, in terms of curves on surfaces. In this note we describe a canonical set of functions we call \emph{surface functions},  associated to all orders in the topological expansion,  that are naturally suggested by this point of view. Surface functions are generating functions for all inequivalent triangulations of the surface. They generalize matrix model correlators, and in the planar limit, coincide with field theoretic loop integrands. We show that surface functions satisfy a universal recursion relation, the \emph{cut equation}, that can be solved without introducing spurious poles, to all orders in the genus expansion. The formalism naturally extends to include triangulations with closed curves, corresponding to theories with uncolored particles. This new recursion is quite different from the topological recursion relations satisfied by  matrix models. Applied to field theory, the new recursion efficiently computes all-order planar integrands for general colored theories, together with uncolored theories at tree-level. As an example we give the all-order recursion for the planar NLSM integrand. We attach a Mathematica notebook for the efficient computation of these planar integrands, with illustrative examples through four loops.}

\begin{document}
  \maketitle

\section{Introduction}

Recent years have seen the emergence of a new way of thinking about the scattering amplitudes for colored particles, at all orders in the topological expansion, associated with a simple counting problem attached to curves on surfaces \cite{counting1,counting2}. The story began with the description of the theory of colored scalars $\phi^i_j$ with cubic interaction Tr $\phi^3$, but has been generalized in a number of ways.  Amplitudes for general colored Lagrangians \cite{tropicalscalars}, couplings to colored fermions \cite{coloredyukawa}, and non-supersymmetric pure Yang-Mills theory can be described in this framework \cite{YM}. And surprisingly, the seemingly toy Tr $\phi^3$ theory turns out to actually contain amplitudes for pions and non-supersymmetric gluons, in any number of dimensions, via a simple kinematic shift \cite{NLSM1}. 

Amongst other things, this picture solves a number of basic, kinematical problems that have bedeviled the very definition of loop integrands using usual momenta.  This is because, in the surface picture, the kinematics is naturally associated not with momenta, but directly with homotopy classes of curves on surfaces. Ordinary loop momenta emerge as a specialization of these variables, in which momenta is determined by homology. 

Even in the planar limit, where conventional notions of loop integrand are well-defined, ``surface kinematics" is more powerful, allowing us to bypass the infamous 1/0 pathologies associated with tadpoles and bubbles on external legs, that are ubiquitous in non-supersymmetric theories \cite{halo,NLSM1}. This makes it possible to define ``perfect" integrands with good properties such as matching all loop-cuts, and enjoying integrand-level gauge invariance for Yang-Mills theory and the Adler zero for the non-linear sigma model \cite{NLSM1,NLSM2,NLSM3}.

For general surfaces, there are infinitely many different curves corresponding to the same propagator, and infinitely many triangulations of the surface corresponding to the same Feynman diagram. This is because of the action of the mapping class group (MCG) on the surface. The MCG also shifts the loop momentum assignments of the propagators. In \cite{counting1}, it was shown how the final, loop integrated amplitude (including beyond the planar limit) can be written as a natural \emph{curve integral} that mods out by the MCG action. However, in the special case of the planar loop integrand, all the curves in the same MCG orbit are assigned the \emph{same} momentum, and hence we can mod out by the MCG already at the level of the integrand.

Motivated by this, we introduce a new natural object that we study in this paper. We will assign kinematic variables to all curves up to homotopy, and identify the variables for curves in the same MCG orbit. For any surface, we can define ``stringy surface functions" via the curve integral
\begin{equation}\label{eq:string}
{\cal G}_S(\alpha^\prime, X_C) = \alpha^{\prime\ {\rm dim}\ T(S)} \int_0^\infty \frac{\omega}{{\rm MCG}} \prod_C u_C^{\alpha^\prime X_C}
\end{equation}
Here we see the $u$-variables, $0 \leq u_C \leq 1$, associated with each curve on the surface, and the canonical form $\omega$, which logarithmic singularities on the boundaries of the \emph{binary geometry} $T(S)$ defined by the $u$-variables \cite{counting1}. We also associate a single kinematic variable, $X_C$, for all curves $C$ in the same MCG orbit. This makes the integrand itself MCG invariant, so it makes sense to mod out by the MCG in the integral, and ensures the stringy surface functions $\mathcal{G}_S$ are well defined.

The stringy surface functions are an interesting generalization of many familiar objects in theoretical physics. In the $\alpha^\prime \to 0$ limit, and setting all $X_C \to 1$, $\mathcal{G}_S$ simply counts all diagrams/triangulations of the surface $S$, and is hence computing contributions to the genus expansion of a matrix model correlator. The functions ${\cal G}_S(\alpha^\prime, X_C)$ are therefore an interesting double-generalization of the matrix model. On the one hand, setting all variables equal, $X_C \to 1$, but keeping finite $\alpha^\prime$, we find an $\alpha^\prime$ deformation of matrix model correlators. On the other hand, keeping the $X_C$ variables distinct, but taking the $\alpha^\prime \rightarrow 0$ limit, we find $\mathcal{G}_S(\alpha', X_C) \to G_S(X_C)$, where $G_S(X_C)$ is a rational function in the $X_C$ that we call a \emph{surface function}. $G_S(X_C)$ can be thought of as the generating function for all MCG-inequivalent triangulations of the surface $S$. Moreover, for a planar surface, $G_S$ is a surface generalization of planar integrands (for cubic $\Tr \phi^3$ theory), that recovers ordinary planar integrands when the $X_C$ are specialized to the appropriate kinematics.

It is convenient to use the inverse variables $x_C \equiv 1/X_C$. Then, for any surface, the surface functions $G_S$ are polynomials in the $x_C$ variables. We find that the surface functions, $G_S$, satisfy a universal recursion, in the form of differential equations that we call the \emph{cut equations}. For a curve $C$ on a surface $S$, the cut equation is
\begin{equation}\label{intro:cut}
\partial_{x_C} G_S = G_{S\setminus{C}},
\end{equation}
where ${S\setminus{C}}$ denotes the surface obtained cutting $S$ along $C$.
Together with boundary conditions at $x_C \to 0$ (which correspond to $X_C \to \infty$, or ``the ultraviolet" in the field-theoretic interpretation), these equations can be integrated to compute surface functions for more complicated surfaces from simpler ones. As we will see, since the $G_S$ are polynomials, this integration is essentially trivial: the cut equation controls the intricate combinatorics of diagrams, and the integrals simply distribute appropriate numerical weights, so that all terms sum together to give the correct result for each surface function. 

The cut equation is very general and holds far beyond triangulations of surfaces. We study a useful family of generalized surface functions that are generating functions for \emph{polyangulations} of surfaces. We can further weight each polygon with a ``kinematic numerator factor'', taken as functions of some kinematic variables $Y_C$, thought of as independent from the $X_C$ that only appear in the denominators of $G_S$. These numerators correspond to the interaction vertices of a colored scalar theory, with a general (single trace) interaction Lagrangian. These generalized surface functions satisfy exactly the same cut equations, \eqref{intro:cut}. This allows us to compute planar integrands using the cut equation for arbitrary colored scalars arbitrary interactions.

We can extend the definition of surface functions even further to capture more general decompositions of surfaces, that also include \emph{closed curves}. Closed curves are those that do not end on boundaries, like the $A-$ or $B-$ cycles on a torus. For field theory, this corresponds to adding uncolored particles, $\sigma$, with arbitrary interactions, and arbitary couplings to colored scalars. The tree amplitudes for the $\sigma$ as well as all-order planar integrands for $\phi$, $\sigma$ scattering (but with no internal $\sigma$ loops) can again be computed using the cut equation.

It is interesting to compare the cut equation with other methods for recursively computing amplitudes. The most familiar way of characterizing amplitudes is via their factorization property on poles, and this is particularly useful to determine tree-level amplitudes. But integrands for non-supersymmetric theories offer a challenge --- the integrands do {\it not} just have simple poles. The presence of bubbles and tadpoles gives rise to a proliferation of higher poles as well. This makes it much more difficult to characterize singularities by residues and use Cauchy-theorem style arguments to determine integrands. The cut equation turns this problem literally on its head. With surface functions, instead of studying $1/X$ singularities in the $X$ variables, we study polynomials in the inverse variables $x =1/X$. The cut equation satisfied by $G_S$ beautifully takes into account the correct combinatorial factors associated with arbitrary powers of $x$. 

In this paper, we begin by defining the surface functions $G_S$ and computing them in examples. We then show the cut equation in action, and show how it can be used to efficiently compute field-theory integrands. A Mathematica package for computing planar integrands of a general colored scalar theory using the cut equation is included for readers interested in working with these objects. In particular, we include results for the non-linear sigma model (NLSM) through to 4-loops. We leave the exploration of surface functions at finite $\alpha^\prime$ to future work.

\section{Surfaces in Perturbation Theory}\label{sec:surface}

\begin{figure}
\begin{center}
\includegraphics[width=0.65\textwidth,trim={.5cm 1cm .5cm 1cm},clip]{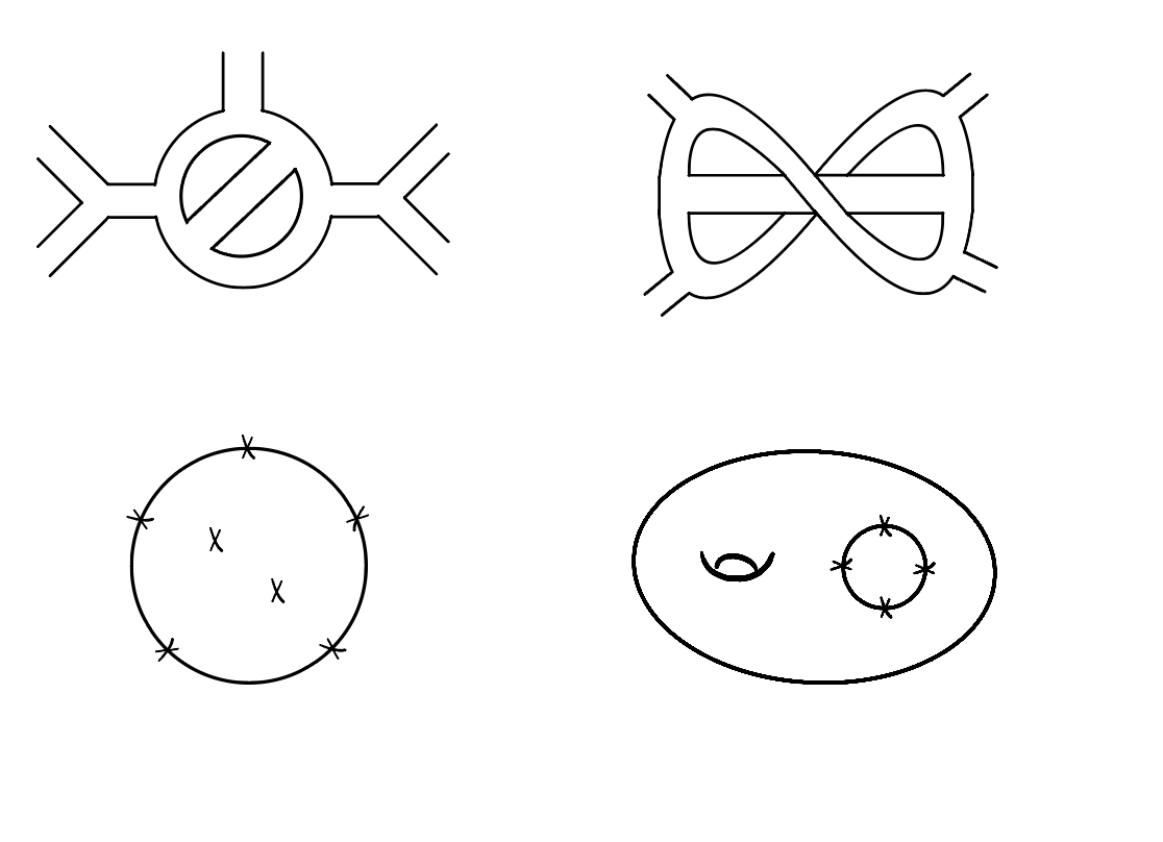}
\caption{Fatgraphs (top row) contribute to partial amplitudes described by surfaces (bottom row). The surfaces capture the color factor of the graph. Factors of $N$ correspond to punctures and trace-factors correspond to boundaries.}
\label{fig:fat}
\end{center}
\end{figure}

Before we define the object of this paper, surface functions, it will be helpful to recall how surfaces arise in ordinary perturbation theory. Consider theories of a single colored scalar field $\phi_I^J$ (with indices valued in the fundamental and anti-fundamental representation of $U(N)$). Any single-trace interaction Lagrangian ${\cal L}_{\rm int}$ for $\phi$ can be written in momentum space, and takes the form
\beq
{\cal L}_{\rm int} = \sum_m \lambda^{m-2} \Tr\left(\phi(k_1) \phi(k_2) \cdots \phi(k_m) \right) \, {\cal L}^{(m)}(k_1,\ldots,k_m) \, \delta \left( \sum_{i=1}^m k_i^\mu \right),
\eeq
for coupling constant $\lambda$. Here, the ${\cal L}^{(m)}$ are some cyclically invariant functions of the momenta defined on the support of the momentum conservation relations, $\sum_{i=1}^m k_i^\mu = 0$. By Lorentz invariance, ${\cal L}^{(m)}$ can be written as a function of the invariants $k_i \cdot k_j$. Equivalently, ${\cal L}^{(m)}$ is a function of the Lorentz invariant variables\footnote{Note that $2k_i \cdot k_j = Y_{i,j+1} + Y_{i+1,j} - Y_{i,j} - Y_{i+1,j+1}$.}
\beq\label{eq:Y}
Y_{i,j} = \left(k_i + k_{i+1} + \cdots + k_{j-1} \right)^2,
\eeq
which correspond to the chords of the $m$-gon formed by the momenta $k_i$. For example, the 4-point vertex $\Tr \left[ (\partial \phi \cdot \partial \phi) \phi \phi \right]/2$ becomes, in momentum space,
\beq
{\cal L}^{(4)}(1234) = Y_{1,3} + Y_{2,4} - \left( Y_{1,2} + Y_{2,3} + Y_{3,4}+Y_{4,1}\right).
\eeq
The Feynman diagrams of the theory are fatgraphs $\Gamma$ with vertices of any degree that appears in ${\cal L}_{\rm int}$. For a fatgraph $\Gamma$ with $e$ internal edges and $v$ vertices, the loop number of the graph is $\ell = e - v + 1$. The boundaries of $\Gamma$ correspond to its trace factor. Each closed loop contributes $\Tr(1) = N$ to the trace factor and a boundary with $m > 0$ external lines contributes a factor for the form $\Tr (t_1 \cdots t_m)$ for some external color matrices $t_i$. If $\Gamma$ has $p$ closed loop boundaries, and $h$ boundary components with external states, then $\Gamma$ can be embedded into a genus $g$ surface with\footnote{This is a consequence of the Euler relation: $2 - 2g = h+p-e+k$.}
\beq\label{eq:euler}
p+h+2g = \ell+1.
\eeq
The color factor of a diagram $\Gamma$ is naturally associated to a surface $S$ with boundaries and we write it as $C_S$: each factor of $N$ in the color factor is associated to an unlabelled puncture on the surface, and each factor of the form $\Tr (t_1 \cdots t_m)$ is associated to a boundary with $m$ marked points. For examples of how the color structure of a diagram $\Gamma$ defines a surface, see Figure \ref{fig:fat}. A \emph{partial amplitude}, $A_S$, is the sum of all diagrams with the same color factor, corresponding to $S$.  Then the $n$-point amplitude has a partial amplitude expansion
\begin{equation}
{\cal A}_{n} = \sum_{\ell=0}^{\infty} \lambda^{n-2+2\ell} \sum_{S}  N^{p} \,A_{S}\, C_S,
\end{equation}
where, for each loop order, $\ell$, we sum over all surfaces $S$ having $p$ punctures, $h$ boundaries and genus $g$, subject to the Euler constraint, \eqref{eq:euler}. It is natural to use the appearance of the surfaces $S$ in the partial amplitude expansion to \emph{compute} the amplitude. A diagram $\Gamma$ that contributes to $A_S$ can be regarded as a collection of \emph{curves} on $S$ that cut $S$ into $m$-gons: each curve contributes a propagator factor $1/X$, and each $m$-gon contributes an interaction vertex ${\cal L}^{(m)}$ that can itself be written in terms of the $X$ variables. One way to compute $A_S$ from the data of the surface is the \emph{curve integral} formalism developed in \cite{counting1,counting2}.

\begin{figure}
\centering
\includegraphics[width=0.65\textwidth,trim={.25cm .1cm .25cm .1cm},clip]{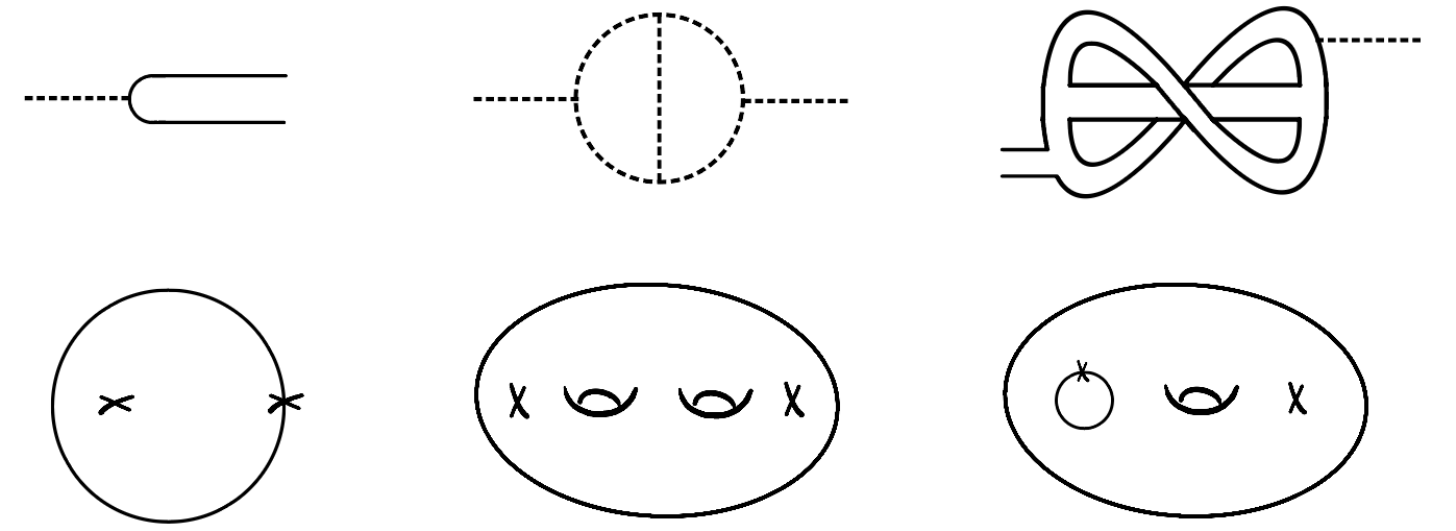}
\caption{The diagrams (top row) of a theory of colored scalars coupled to an uncolored scalar are organized into partial amplitudes labelled by surfaces (bottom row). Here, dashed lines are uncolored scalar propagators.}
\label{fig:thinfat}
\end{figure}

We also consider theories that include an uncolored scalar field, $\sigma$. A term in the ${\cal L}_{\rm int}$ for such a theory takes the form
\beq
\Tr\left(\phi(k_1)\cdots \phi(k_a)\right)\, \sigma(p_1)\cdots \sigma(p_{b})\, \times \, {\cal L}^{(a,b)} (k_1,\ldots,k_a; p_1,\ldots,p_b),
\eeq
for some function ${\cal L}^{(a,b)}$ that is cyclically symmetric in the $k_i$, permutation invariant in the $p_i$, and with support on the momentum conservation relation ($\sum_i k_i + \sum_j p_j = 0$). Moreover, ${\cal L}^{(a,b)}$ is necessarily a function of Lorentz invariants. Each ${\cal L}^{(a,b)}$ can be expressed as a function of the $Y_{i,j}$ as well as the invariants
\beq
Y_A = \left( \sum_{i \in A} p_i \right)^2
\eeq
for subsets $A \subset \{1,\ldots,b\}$ of the $\sigma$ state momenta.\footnote{If $p_i \cdot k_j$ appears in the interaction, it must, by cyclic invariance, appear as $\sum_j p_i \cdot k_j$. But by momentum conservation $\sum_j k_j = - \sum_i p_i$. So in fact these contributions to the interaction can be written as functions of the $Y_A$ invariants.} For example, the vertex $(\partial \sigma \cdot \partial \sigma) \sigma^2/2$ becomes, in momentum space,
\beq
{\cal L}^{(4)}(1234) = Y_{12} + Y_{23} + Y_{13} - 2( Y_1 + Y_2 + Y_3 + Y_4),
\eeq
and the vertex $\sigma \partial_\mu \sigma \Tr \left[ \phi \partial^\mu \phi \right]$ becomes
\beq
{\cal L}^{(2,2)}(12;34)= - (p_3+p_4)^2 = - Y_{34},
\eeq
where $p_3, p_4$ are the momenta of the $\sigma$ states.

A Feynman diagram $\Gamma$ for this theory is a graph that has both fat edges ($\phi$ propagators) and ordinary edges ($\sigma$ propagators). If $\Gamma$ has $e$ internal edges and $v$ vertices, it appears at loop order $\ell = e - v + 1$. It is again natural to organise these contributions into partial amplitudes $A_S$ for surfaces $S$ with punctures and boundaries, as illustrated for some examples in Figure \ref{fig:thinfat}.

Finally, a diagram $\Gamma$ contributing to $A_S$ is dual to a collection of curves on $S$. $\phi$ propagators are dually \emph{open curves}, that begin and end on boundaries and punctures, and $\sigma$ propagators are dual to \emph{closed curves}. The curves cut $S$ into simple surfaces that correspond to the interaction vertices.\footnote{$m$-gons are associated to colored vertices ${\cal L}^{(m,0)}$; $m$-gons with $b$ punctures associated to vertices ${\cal L}^{(m,b)}$; and spheres with $b$ punctures associated to the vertices ${\cal L}^{(0,b)}$ that involve only uncolored $\sigma$ states.} However, instead of further studying the full perturbation series, we now turn to the object of this paper: a simpler family of functions, \emph{surface functions}, inspired by how surfaces appear in the perturbation series.

\section{Surface Functions}
For any fixed theory ${\cal L}_{\rm int}$, there is a family of \emph{surface functions} $G_S$, which are a unique rational function for each marked surface $S$. They coincide with integrands for the theory in the planar limit (i.e. when $S$ is a punctured disk), but are defined for all surfaces. Moreover, $G_S$ can be regarded as a deformation of contributions to certain matrix model correlators (see Section \ref{sec:matrix}). We will first study surface functions for theories of a colored scalar $\phi$. See Section \ref{sec:uncolored} for theories with an uncolored scalar field.

We begin, in this section, with the simple case of ${\cal L}_{\rm int} = \Tr ( \phi^3) / 3$, for a single colored scalar $\phi$. We define the surface functions $G_S$ for this theory to be generating series that record all possible inequivalent triangulations of $S$. Two triangulations are \emph{equivalent} if they are related by the action of the Mapping Class Group, $\MCG$ --- in other words, they are equivalent if they define the same fatgraph $\Gamma$. Introduce variables $x_C$ for each $\MCG$-class of curves $C$ on the surface. Then $G_S$ is the sum\footnote{If a $\Gamma$ in this sum has nontrivial automorphisms, we divide its term by the symmetry factor ${\rm Aut}_\Gamma$ of that graph. See section \ref{sec:symmetry}.}
\beq\label{eq:Gdef}
G_S = \sum_\Gamma \prod_{C\in \Gamma} x_C
\eeq
over all inequivalent triangulations $\Gamma$ of $S$.
Although surface functions can be defined for any surface $S$ we are mostly interested in the case where $S$ is a planar surface: a disk with a set of labeled marks, $A$, and a set of punctures, $B$, in the bulk. 
Accordingly we also write $G(A;B)$ for the corresponding surface function.

For example, if $S$ is a disk with $n$ marked points, $G_S$ is the sum over the Catalan-many triangulations of the disk, and we write this as
\beq\label{eg:phi3:tree}
G(1234) = x_{13} + x_{24},\qquad G(12345) = x_{13}x_{14} + ({\rm cyclic}\, 12345),
\eeq
where $x_{ij}$ is the curve from point $i$ to point $j$. The tree amplitudes are recovered by substituting the kinematic data as $x_{ij} \rightarrow 1/X_{ij}$, where $X_{ij} = (k_i + \cdots + k_{j-1})^2 + m^2$ are the propagator factors.

\begin{figure}
\centering
\includegraphics[width=0.65\textwidth,trim={.25cm .1cm .25cm .1cm},clip]{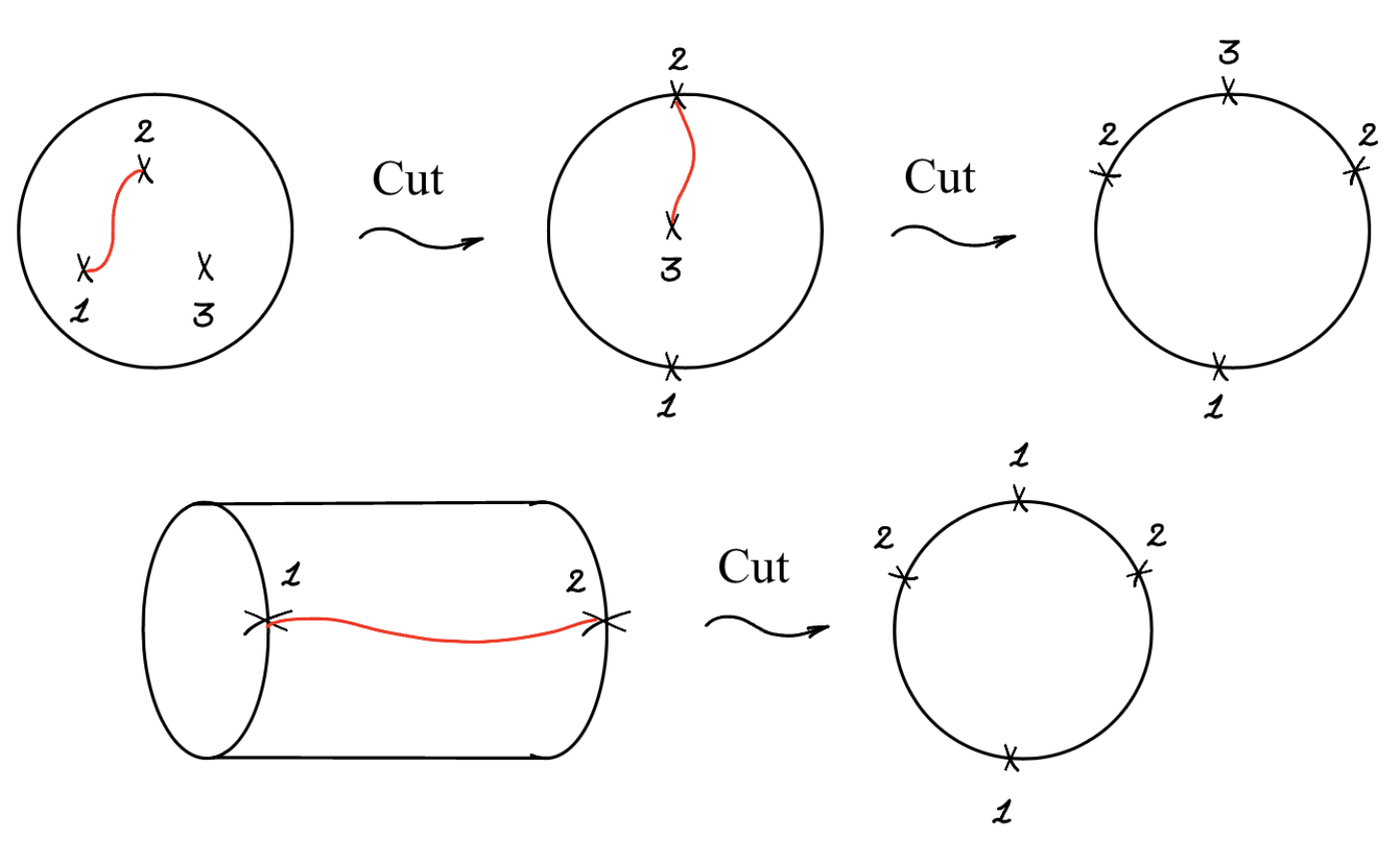}
\caption{The surface functions capture the cutting of a surface along curves to get simpler surfaces. The figure illustrates the three cuts discussed in the text (see equations \eqref{eg:phi3:cut1}, \eqref{eg:phi3:cut2}, and \eqref{eg:phi3:cut3}, respectively).}
\label{fig:cut}
\end{figure}

At 1-loop, consider, for example, the disk with two marked points, labelled $1$ and $2$, and one puncture, labelled $3$. For this surface there are four variables: $x_{13}$, $x_{23}$, $x_{11}$, and $x_{22}$, where $x_{ij}$ is the curve with endpoints $i$ and $j$. The surface function for this surface is then
\beq\label{eg:phi3:n2l1}
G(12;3) = x_{13}x_{11} + x_{13}x_{23} + x_{23}x_{22}.
\eeq
The terms correspond to the two tadpole and one bubble diagram. An integrand for the 1-loop propagator is recovered from $G(12;3)$ by substituting $x_C$ with the propagator factors $1/(P_C^2 + m^2)$, for some choice of loop momentum routing.\footnote{There is a canonical way to associate curves with momenta, see \cite{counting1}. However, we do not study loop integration here.} We can already see in this example how the surface function captures the cuts of the loop integrand, and how this is related to the geometry of the surface. For example, taking a derivative of \eqref{eg:phi3:n2l1},
\beq\label{eg:phi3:cut1}
\partial_{x_{23}} G(12;3) = x_{22}+x_{13} = G(1232),
\eeq
we recover a 4-point tree amplitude, \eqref{eg:phi3:tree}. Indeed, cutting the punctured disk $S(12;3)$ along $x_{13}$ gives the 4-point disk with marked points $1213$ (see Figure \ref{fig:cut}). This is an example of the \emph{cut equation} studied in the next section.

For a 2-loop example, consider a sphere $S$ with three punctures, labelled $1$, $2$, $3$. Write $G(\emptyset; 1;2;3)$ for its surface function, which corresponds to the 2-loop planar contribution to the vacuum partition function. There are three variables $x_{ij}$ for the curves connecting distinct punctures, and three variables $x_{ii}$ for curves that begin and end at the same puncture. We find
\beq\label{eg:phi3:n0l2}
G(\emptyset;\,1,2,3) = x_{12}x_{23}x_{13} + \left[ x_{11}x_{12}x_{13} + x_{22} x_{12}x_{23} + x_{33} x_{13}x_{23}\right].
\eeq
Once again, an integrand for this contribution to the vacuum partition function can be obtained by substituting the $x$ variables with propagator factors: each term corresponds to one of the 5 planar vacuum 2-loop graphs. Moreover, the derivatives of this surface function compute the appropriate cuts. For example,
\beq\label{eg:phi3:cut2}
\partial_{x_{12}} G(\emptyset;\,1,2,3) = x_{23}x_{13} + x_{13}x_{11} + x_{23}x_{22} = G(12;3),
\eeq
which agrees with the fact that cutting the sphere along $x_{12}$ gives a 2-point disk with one puncture (see Figure \ref{fig:cut}).

As a final illustrative example, take $S$ to be the annulus with one marked point on each boundary, labelled $1$ and $2$. There is only one variable, $x_{12}$, corresponding to the curves connecting $1$ and $2$ (which form one $\MCG$ class). Then our definition, \eqref{eq:Gdef}, gives
\beq\label{eg:phi3:annulus}
G_{\rm annulus} = x_{12}^2,
\eeq
since there is only one possible triangulation/fatgraph, $\Gamma$. Substituting a propagator factor for the variable $x_{12}$ does \emph{not} recover an integrand for the non-planar amplitude for $\Tr \phi^3$ theory. For the non-planar integrand, the propagators of $\Gamma$ must be assigned \emph{different} momenta, to recover the standard Feynman integral
\beq
A_{\rm annulus} = \int d^D\ell \, \frac{1}{\ell^2+m^2} \frac{1}{(\ell + k)^2 + m^2},
\eeq
where $k^\mu$ is the external momentum. In general, surface functions do not compute integrands for the theory beyond the planar limit. However, they continue to capture the factorisation properties of the surface. Indeed,
\beq\label{eg:phi3:cut3}
\partial_{x_{12}} G_{\rm annulus} = x_{12} + x_{12} = G(1212),
\eeq
which is precisely the surface function of the 4-point disk that is obtained by cutting the annulus along the curve $x_{12}$ (see Figure \ref{fig:cut}).

\section{Cut Equation}\label{sec:cut}
In the examples above, equations (\ref{eg:phi3:cut1}--\ref{eg:phi3:cut3}), we have seen that the $\Tr \phi^3$ surface functions satisfy the \emph{cut equation}
\beq\label{eq:cut}
\partial_{x_i} \, G_{S}(x_1,x_2,\ldots) = G_{S\,\text{cut along}\,x_i}(x_1,x_2,\ldots),
\eeq
where $x_i$ is one of the variables appearing in $G_S$, corresponding to a $\MCG$ class of curves on $S$. In this section we explain why the cut equation holds for the $\Tr \phi^3$ surface functions. However, it is also natural to instead \emph{define} surface functions as solutions of the cut equation. We will do this systematically in the next section by solving the cut equation recursively.

\subsection{Origin of the Cut Equation}\label{sec:cut:origin}
The cut equation follows from the definition \eqref{eq:Gdef} of the $\Tr \phi^3$ surface functions, and we will consider it case by case.

In the simplest case, suppose $G_S$ is linear in $x_i$. Then the surface function has the form $G_S = A + B x_i$. $A$ is the sum over all $\Gamma$ that do not involve $x_i$. Whereas $B$ is a sum over all triangulations that involve $x_i$: but this is the same thing as summing over all triangulations of the surface obtained by cutting $S$ along $x_i$. So $B = G_{S\,{\rm cut}}$ and the cut equation \eqref{eq:cut} holds in this case. 

As a special case, $G_S$ is always linear in $x_i$ if $x_i$ is a curve that \emph{factorizes} $S$ into two separate surfaces, $S_L$ and $S_R$. In this case we have
\beq
G_S = A +  G_{S_L} G_{S_R} x_i,
\eeq
which implies the cut equation
\beq
\partial_{x_i} G_S = G_{S_L} G_{S_R}
\eeq
for the case when $x_i$ factorizes $S$.

\begin{figure}
\centering
\includegraphics[width=0.75\textwidth,trim={.25cm .1cm .25cm .1cm},clip]{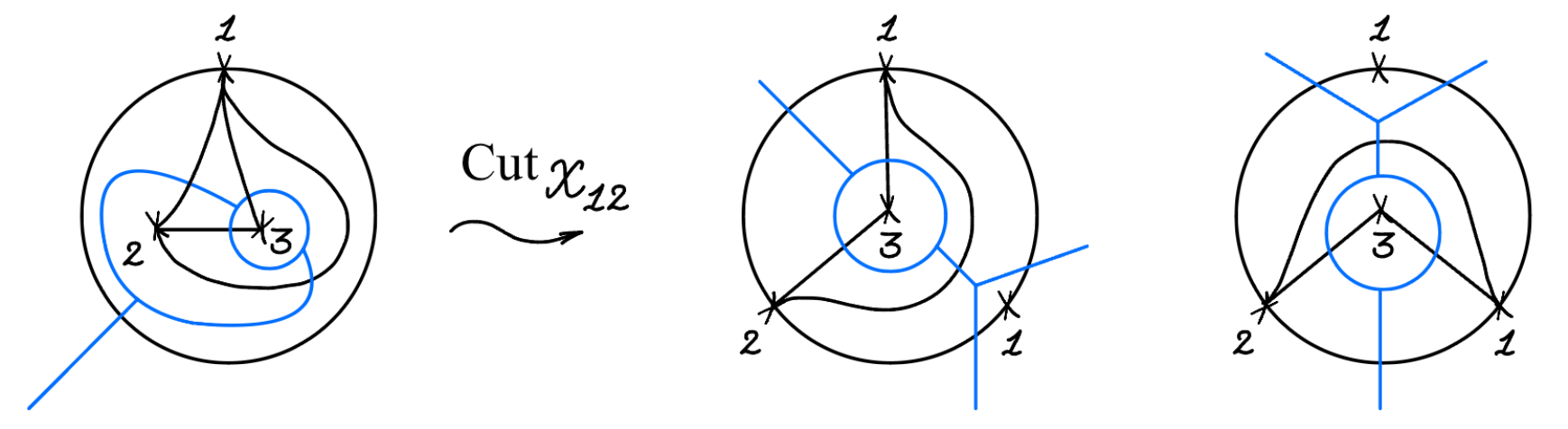}
\caption{Cutting along a class of curves that appear multiple times in a single triangulation (black lines) or diagram (blue lines) gives distinct triangulations of the cut surface. The triangulation on the left corresponds to the term $x_{12}^2x_{13}x_{23}$ in $G(1;23)$. The two distinct triangulations on the right both contribute $x_{12}x_{13}x_{23}$ to $G(121;3)$. The cut equation automatically takes care of the factor of $2$ that appears in this way.}
\label{fig:bubble}
\end{figure}

Finally, suppose $G_S$ is not linear in $x_i$. This can happen if $x_i$ is a curve that does not cut $S$ into two separate surfaces. Consider a term in $G_S$, $x_i^k F$, for $k>1$ and some monomial $F$ in the other variables. This term, $x_i^k F$,  corresponds to some triangulation of $S$: the $k$ copies of $x_i$ correspond to $k$ distinct curves on $S$ (i.e. they are not homotopic, but are $\MCG$ equivalent). We can choose to cut along any one of these $k$ copies of $x_i$ to get a triangulation on the cut surface $S_{\rm cut}$. However, each of these $k$ triangulations is  \emph{inequivalent} on $S_{\rm cut}$, because the mapping class group of $S_{\rm cut}$ is  the stabilizer subgroup ${\rm Stab}(x_i) \leq \MCG$ of the original $\MCG$ of $S$. In other words, on $S_{\rm cut}$, the $k$ copies of $x_i$ can no longer be mapped to each other under the action of the new $\MCG$. And so, cutting along $x_i$, this monomial in $G_S$ contributes
\beq
x_i^k F \in G_S ~~ \longrightarrow ~~ k x_i^{k-1} F \in G_{S\,{\rm cut}}
\eeq
to the surface function of $S_{\rm cut}$. But this is the same as taking the derivative with respect to $x_i$. So this implies the cut equation \eqref{eq:cut}.

As an illustrative example of this last case, consider the 1-loop 2-point surface function $G(1;23)$, with $1$ on the boundary, and punctures labelled $2,3$. This contains the term, e.g., $x_{12}^2 x_{13}x_{23}$, which corresponds to a diagram with an internal bubble. Cutting the surface along $x_{12}$ gives the 1-loop disk $G(121;3)$. And we check that cutting the triangulation corresponding to $x_{12}^2 x_{13}x_{23}$ along either copy of $x_{12}$ gives \emph{two} distinct 1-loop bubble triangulations, but both with the three curves $x_{12}x_{13}x_{23}$. So that the term $x_{12}^2 x_{13}x_{23} \in G(1;23)$ becomes the term $2 x_{12}x_{13} x_{23}$ in $G(121;3)$. See Figure \ref{fig:bubble}.

The above orbit-stabilizer type argument is very natural at the level of string- and curve-integrals, like \eqref{eq:string}, which also give a further motivation for studying these surface functions \cite{counting1,counting2, YM, YM2}. See Appendix \ref{app:ap} to see how the cut equation is implied by these these integrals.

\subsection{Symmetry factors}\label{sec:symmetry}
One consequence of the cut equation is that we see why it is natural to define surface functions $G_S$ as a sum over diagrams weighted by symmetry factors. Consider the torus with one puncture, labelled $1$. Its surface function in $\Tr \phi^3$ theory, $G_{\rm torus}$, is function of only one variable, $x_{11}$, since there is only one class of curves on this surface. It should satisfy the cut equation
\beq
\partial_{x_{11}} G_{\rm torus} = G_{\rm annulus}.
\eeq
But $G_{\rm annulus} = x_{11}^2$ (see equation \eqref{eg:phi3:annulus}). So we learn that we should define $G_{\rm torus}$ as
\beq
G_{\rm torus} = \frac{1}{3} x_{11}^3.
\eeq
This factor of $1/3$ is precisely the symmetry factor, ${\rm Aut}_\Gamma$, of the fatgraph, $\Gamma$, that contributes to the vacuum at genus-one. So the surface functions $G_S$ for $\Tr \phi^3$ theory should be defined as sums over distinct fatgraphs weighted by symmetry factors:
\beq\label{eq:Gdefaut}
G_S = \sum_\Gamma \frac{1}{{\rm Aut}_\Gamma} \prod_{C\in \Gamma} x_C.
\eeq
However, note all non-vacuum fatgraphs have ${\rm Aut}_\Gamma = 1$, so in every case (except vacuum) we recover the original definition given in \eqref{eq:Gdef}.

\section{Surface Recursion}
In this section we solve the cut equation using a simple recursion. This recursion naturally produces surface functions for all colored scalar theories ${\cal L}_{\rm int}$. In particular, we apply this to computing the planar integrands of the non-linear sigma model (NLSM) in Section \ref{sec:NLSM}.

Take a surface function $G_S$, which is a polynomial in some variables, $x_i$. Now rescale some subset, $\mathcal{S}$, of these variables as $x_i \mapsto t x_i$ for all $i$ in the set $\mathcal{S}$, so that $G_S$ becomes a function of $t$. Then, by the cut equation,
\beq\label{eq:dert}
\partial_t G_S(t)  = \sum_{i\in \mathcal{S}} x_i \, G^\text{cut $i$}_S(t),
\eeq
where $G^\text{cut $i$}_S(t)$ is the surface function for the surface obtained by cutting $S$ along $x_i$. It follows that $G_S$ can be recursively solved from simpler functions via
\beq\label{eq:recursion}
G_S = G_S(0) + \int_0^1 dt\, \sum_{i\in \mathcal{S}} x_i \, G^\text{cut $i$}_S(t),
\eeq
for some boundary term $G_S(0)$. Note that $G^\text{cut $i$}_S(t)$ is just a polynomial in $t$, so the $t$-integral just weights different monomials by simple factors, $1/k$, where $k$ is the number of shifted variables in the monomial. The term $G_S(0)$ includes any terms that do not include any of the variables in shift set ${\cal S}$. It is therefore natural to look for shift sets $\mathcal{S}$ such that $G_S(0)$ is as simple as possible.

The surface functions for $\Tr\phi^3$ theory can be reproduced using this recursion, \eqref{eq:recursion}, by choosing an appropriate shift set, ${\cal S}$. We will choose ${\cal S}$ so that every term in $G_S$ contains at least one variable from ${\cal S}$, which means that, at $t=0$,
\beq
G_S (0) = 0.
\eeq
One option is to shift \emph{all} variables. However, there are many other choices.

For example, for the 5-point tree-level surface function $G(12345)$ (equation \eqref{eg:phi3:tree}), one possibility is to shift the three variables $x_{13}, x_{14}, x_{25}$. This works because every diagram/term contributing to $G(12345)$ must contain at least one of these 3 variables. So, using \eqref{eq:recursion}
\begin{align}
G(12345) & = \int_0^1 dt\left[ x_{13} G(123) G(3451) + x_{14} G(1234) G(451) + x_{25} G(2345) G(512)\right]\\
& = \int_0^1 dt\left[ x_{13} (x_{35} + x_{14} t) + x_{14} (x_{13}t + x_{24}) + x_{25} (x_{24} + x_{35})\right]\\
& = x_{13} x_{35} + \frac{1}{2} x_{13}x_{14} + \frac{1}{2} x_{13}x_{14} + x_{14} x_{24} + x_{24} x_{25} + x_{25} x_{35}.
\end{align}
Note how the factors of $1/2$ produced by the $t$-integral ensure that we do not overcount the $x_{13}x_{14}$ term, and correctly recover the expression in \eqref{eg:phi3:tree}. For higher-point tree-level computations we can take the shift set comprising $x_{2n}$ and $x_{1i}$ (for $i=3,\cdots,n-1$). Alternatively the shift set comprising just the short chords, $x_{i,i+2}$, also works well. See Appendix \ref{app:tree} for more details about implementing surface recursion for tree amplitudes and a comparison to Berends-Giele-like recursions.

For planar surface functions at 1-loop and higher, we can implement the recursion using a \emph{simple planar shift} by shifting just those variables $x_{ip}$ that connect a boundary point $i$ to a puncture $p$. For example, the 1-loop 2-point function $G(12;3)$ given in \eqref{eg:phi3:n2l1} can be computed by shifting $x_{13}$, $x_{23}$. Then, using \eqref{eq:recursion},
\begin{align}
G(12;3) & = \int_0^1 dt\, x_{13} G(1312)(t) + x_{23} G(1232)(t) \notag \\
& = \int_0^1 dt\, x_{13} (x_{11} + x_{23} t) + x_{23} (x_{13} t + x_{22})  \label{eg:recursion123} \\
& = x_{13} x_{11} + \frac{1}{2} x_{13} x_{23}  + \frac{1}{2} x_{23} x_{13}  + x_{23} x_{22} \notag
\end{align}
Once again the factors of $1/2$ correctly ensure that we do not overcount. All planar integrands for $\Tr \phi^3$ theory can be obtained in this way. See Appendix \ref{app:code} for further comments about implementing the recursion in practice.

\subsection{Arbitrary Colored Scalar Theories}
However, if we instead try to solve \eqref{eq:recursion} with arbitrary boundary terms $G_S(0)$, we can recover a family of surface functions $G_S$ for every colored scalar theory ${\cal L}_{\rm int}$. Recall from Section \ref{sec:surface} that ${\cal L}_{\rm int}$ can be written in momentum space. Each $m$-point term in the Lagrangian, ${\cal L}^{(m)}(12\cdots m)$, is some polynomial function of the momentum invariants $Y_{i,j}$ (equation \eqref{eq:Y}).

For a simple motivating example, consider the 4-point tree-level surface function $G(1234)$. For $\Tr \phi^3$ theory this is given in \eqref{eg:phi3:tree}. We can take our shift set to be all of the variables $x_{13}$, $x_{24}$, corresponding to the two chords of the disk. Then
\beq
G(1234) = G(1234)(0) + \int_0^1 dt\, x_{13} G(123) G(341) + x_{24} G(234) G(412).
\eeq
If we set $G(1234)(0) = 0$ and $G(123) = 1$, then we recover $G(1234) = x_{13} + x_{24}$ (equation  \eqref{eg:phi3:tree}). However, we are free to make other choices. Suppose we care about a colored scalar theory with arbitrary interactions, ${\cal L}_{\rm int}$. Associated to this Lagrangian, we find a new solution to the cut equations by taking
\beq
G(123) = {\cal L}^{(3)}(123),\qquad G(1234)(0) = {\cal L}^{(4)}(1234).
\eeq
This gives
\beq
G(1234) ={\cal L}^{(4)}(1234) + x_{13} {\cal L}^{(3)}(123){\cal L}^{(3)}(341) + x_{24} {\cal L}^{(3)}(234){\cal L}^{(3)}(431).
\eeq
We can regard this as a polynomial in the variables $x_{i,j}$ and $Y_{i,j}$. Upon substituting the kinematic data, $x_{13} \rightarrow 1/((k_1+k_2)^2+m^2)$, $x_{24} \rightarrow 1/((k_2+k_3)^2+m^2)$, and $Y_{i,j} \rightarrow (k_i + \cdots k_{j-1})^2$, $G(1234)$ reproduces the 4-point amplitude for this theory.

For a given Lagrangian ${\cal L}_{\rm int}$, define the surface functions $G_S$ for this theory as the solution to the cut equations with boundary conditions given by the interaction terms in the Lagrangian. For the $m$-point tree-level disks, we specify that
\beq
\left. G(12\cdots m) \right|_{x\rightarrow 0} = {\cal L}^{(m)}(12\cdots m)
\eeq
when all variables $x$ are sent to $0$. Whereas for higher-loop surfaces, we demand that
\beq
\left. G_S \right|_{x\rightarrow 0} = 0.
\eeq
By surface recursion, these boundary conditions uniquely specify surface functions $G_S$ for all surfaces. 
We can alternatively define $G_S$ as a natural generalization of \eqref{eq:Gdef},
\begin{align}
G_S = \sum_\Gamma \left( \prod_{C \in \Gamma} x_C \right) N_\Gamma(Y_C),
\label{eq:GGsurfdef}
\end{align}
where we sum over all \emph{polyangulations}, $\Gamma$, of $S$ (up to $\MCG$) that are built only out of $m$-gons for $m$ such that $\Lint$ has a nonzero $m$-valent interaction, ${\cal L}^{(m)} \neq 0$. For such a polyangulation, the numerator factor $N_\Gamma(Y_C)$ is the product over the vertices ${\cal L}^{(m)}(Y_C)$ for each $m$-gon appearing in $\Gamma$. 
It is easy to show that \eqref{eq:GGsurfdef} is equivalent to defining $G_S$ using the surface recursion with boundary conditions. We stress that it is important to treat the $Y_C$ variables in the numerators as independent from the $x_C$ variables.

Equation \eqref{eq:GGsurfdef} makes clear that it is possible to recover the planar loop integrands of a general colored theory from the generalized surface functions $G_S$ for planar surfaces. The integrand is obtained from $G_S$ by substituting kinematics: $Y_C \rightarrow P_C^2$ and $X_C \rightarrow P_C^2 + m^2_C$. The planar integrands for an arbitrary colored scalar theory can therefore be computed using surface recursion. This is implemented the included Mathematica notebook.
See also Appendix \ref{app:code} for some further comments about implementing the recursion for planar integrands.

\subsection{NLSM}\label{sec:NLSM}
As a particularly interesting application of surface recursion, we now consider applying it to computing planar integrands for the non-linear sigma model (NLSM). Adopting the minimal parametrization Lagrangian of \cite{JaraNLSM}, the even-point interaction are, in momentum space,
\beq
{\cal L}^{(2n)} = \frac{1}{2^{n-1}} \, \sum_{k=0}^{n-2} c_k \, c_{n-2-k} \, \sum_i Y_{i,i+2k+2},
\eeq
\beq
{\cal L}^{(2n)} = \frac{(-1)^n}{2} \, \sum_{k=0}^{n-2} c_k \, c_{n-2-k} \, \sum_i Y_{i,i+2k+2},
\eeq
where $c_k = {2k\choose k} \frac{1}{k+1}$ are the Catalan numbers. For example,
\beq
{\cal L}^{(4)}(1234) =  (Y_{1,3}+Y_{2,4} ),\qquad {\cal L}^{(6)}(123456) = \frac{1}{2}(Y_{1,3} + Y_{2,4}+Y_{3,5}+Y_{4,6}+Y_{5,1}+Y_{6,2}).
\eeq
Due to the vanishing of the odd vertices, the first amplitude occurring at tree level involves four particles, and it is purely the contact term from the Lagrangian:
\begin{align}
    G(1234) = Y_{1,3} + Y_{2,4}.
\end{align}
Surface recursion can be applied to find higher-point tree amplitudes. For example, at six-points, we have contributions from the factorizations into two 4-point amplitudes. The result is
\begin{align}
    G(123456) = x_{1,4} G(1234)G(4561) + (\mathrm{cyclic}) + \mathcal{L}^{(6)}(123456).
\end{align}

Surface recursion becomes more interesting and useful at loop level. The first non-vanishing integrand is the one loop propagator, $G(12;p)$, where we label the puncture as $p$. Surface recursion computes this as
\begin{align}
G(12; p) &= \int_0^1 dt\ x_{1,p} G(1p12)(t) + \int_0^1 dt\ x_{2,p} G(12p2)(t) 
\notag \\
& = x_{1,p} (Y_{1,1} + Y_{2,p}) + x_{2,p} (Y_{2,2} + Y_{1,p}).\label{eq:NLSMG12p}
\end{align}
We recover an integrand from this surface function, \eqref{eq:NLSMG12p}, by replacing the $x$ variables with the associated propagators, $1/X$, and noticing that, for a massless theory, $X_{i,j} = Y_{i,j}$. Then we obtain the 1-loop integrand
\beq\label{eq:NLSM1}
G(12; p) \rightarrow \frac{X_{1,1}+X_{2,p}}{X_{1,p}} + \frac{X_{2,2}+X_{1,p}}{X_{2,p}}.
\eeq
Similarly, using surface recursion, we also obtain integrands for the 4-point 1-loop integrand,
\begin{align}\label{eq:NLSM2}
    G(1234;p) \rightarrow  \frac{\left(X_{2,{p}} + X_{1,3}\right)\left(X_{4,p}+X_{1,3}\right)}{X_{1,p}X_{3,p}}+\frac{\left(X_{1,{p}} + X_{2,4}\right)\left(X_{3,p}+X_{2,4}\right)}{X_{2,p}X_{4,p}}+(\rm{scaleless}),
\end{align}
and the 2-loop propagator,
\begin{align}\label{eq:NLSM3}
    G(12;p_1,p_2) = \frac{(X_{1,p_2}+X_{2,p_1})^2}{X_{1,z1}X_{2,p_2}X_{p_1,p_2}} + (p_1 \leftrightarrow p_2) + (\rm{scaleless}).
\end{align}
Here, ``scaleless'' denotes a sum of terms that that define scaleless integrals, which integrate to zero in dimensional regularization.

The ancillary Mathematica notebook contains the complete results for NLSM integrands computed using surface recursion, including the scaleless terms, as well as new results up to four loops. These can be independently reproduced with the provided code, in a few minutes on an ordinary laptop. See also Appendix \ref{app:code} for further remarks on implementing surface recursion for planar integrands in practice.

It is interesting to compare the above results with a different description for NLSM integrands, which was introduced in \cite{NLSM1,NLSM2,NLSM3}. There the NLSM integrands are extracted from a suitable shift of $\mathrm{Tr}(\phi^3)$ integrands, which themselves were calculated from the recursion described in this paper. We verify in all examples we have computed, including (\ref{eq:NLSM1}--\ref{eq:NLSM3}), that surface recursion agrees with the results of \cite{NLSM1,NLSM2,NLSM3} up to scaleless integrals. However, a direct comparison shows that we obtain different functions. For instance, our $G(12;p)$ differs from theirs by
\begin{align}\label{eq:NLSMshift}
    G(12;p)^{\text{from shift}}-G(12;p)=2.
\end{align}
Naively the two results seem incompatible, but the mismatch of $2$ is a scaleless integrand, which integrates to zero in dimensional regularization. 

The possibility of adding scaleless integrals introduces an interesting ambiguity in the definition of loop integrands. This is a new ambiguity, in addition to the usual ambiguity that arises from choosing loop momenta. The integrand proposed in \cite{NLSM1,NLSM2,NLSM3} resolves this scaleless ambiguity by making a canonical choice: it is extracted from the low energy limit of a natural $\delta$ deformation of the Tr $\phi^3$ integrand. Most strikingly, the integrand posses the Adler zero, which is usually only expected at the integrated amplitude level. (See also \cite{Bern:2024vqs} for work on finding integrands for NLSM amplitudes.)

From the above examples, it seems plausible that we can compute the canonical NLSM integrands of \cite{NLSM1,NLSM2,NLSM3} at all orders \emph{using} surface recursion, by adding some local boundary terms (i.e. no propagators) to the recursion. In \eqref{eq:NLSMshift} above, the extra term of $2$ can be included in the recursion as a boundary term, $G(12;p)(0)$, associated to the 2-point punctured disk. This raises the interesting question of whether we can compute ``canonical'' integrands of other theories by imposing a set of prescribed zeroes, and including appropriate boundary conditions in the surface recursion.

\subsection{Surface Recursion vs. Cauchy Residue Theorem}\label{sec:cauchy}
Before going further, it is instructive to contrast surface recursion with BCFW-like recursions that compute integrands using the Cauchy residue theorem. Both methods use `shift sets' and reproduce integrands from simpler (lower-point, lower-loop) integrands. There are three main differences.  First, tadpole graphs in BCFW-like recursions have to accounted for separately, unlike in surface recursion.  Second, surface recursion does not produce spurious poles. Third, surface recursion admits a larger class of possible shift sets. 

We can already illustrate each of these three differences with the simple case of 1-loop planar integrands for, say, $\Tr \phi^3$ theory. Write $G(1\cdots n; 0)$ for the $n$-point planar integrand. We can view this as a function of the $x$ variables, or as a function of their inverses, $X = 1/ x$. 

\paragraph{Tadpoles.} To compute $G(1\cdots n; 0)$ using \emph{either} surface recursion or the Cauchy theorem, the shift set ${\cal S}$ should be chosen so that every term in $G$ has at least one propagator from ${\cal S}$. For example, we can take the variables $X_{0i}$ connecting the puncture $0$ to each of the external particles $i = 1,\ldots,n$. For this set, we can try to proceed with a BCFW-like recursion. Shifting each of the $X_{0i}$ by the BCFW-like shift $X \rightarrow X + z$. This gives a function $G(z)/z$ with simple poles at $z = - X_{0i}$. However, this function does not vanish at $\infty$. For large $z$,
\begin{equation}
\frac{G(z)}{z} \sim \frac{1}{z^2} ({\rm tadpoles}) + O(z^{-3}).
\end{equation}
So the usual residue theorem argument can only be applied if we ignore tadpole propagators by setting $1/X_{\rm tadpole} \mapsto 0$. By contrast, we are free to keep tadpole propagators in surface recursion.

\paragraph{Spurious poles.} The Cauchy residue theorem applied to $G(z)/z$ gives
\begin{equation}\label{eq:BCFWsum}
G(0) = \sum_i \frac{1}{X_{0i}} {\rm Res}_{z=-X_{0i}} G(z).
\end{equation}
Each of these residues corresponds to some cut of the 1-loop integrand. Indeed, because there are no double poles in the $X_{0i}$, this can be rewritten as
\begin{equation}
G = - \sum_{i=1}^n X_{0i} \left(G_{{\rm cut}\, X_{0i}}(X)\right)_{X_{0j}\rightarrow X_{0j}-X_{0i}}.
\end{equation}
Note that this appears to be similar to the surface recursion, which reads
\beq
G = \int_0^1 dt\, \sum_{i=1}^n \frac{1}{X_{0i}} \left(G_{{\rm cut}\, X_{0i}}(X)\right)_{X_{0j}\rightarrow X_{0j}/t},
\eeq
where we have written this in terms of the $X$ variables rather than the $x = 1/X$ variables. However, in the BCFW-like recursion, the terms in \eqref{eq:BCFWsum} have spurious poles of the form $1/(X_{0i} - X_{0j})$. For example, at 3-points, the BCFW-like recursion gives
\begin{multline}\label{eq:BCFW31}
G = \frac{1}{X_{01}} \left( \frac{1}{(X_{02}-X_{01})(X_{03}-X_{01})} + \frac{1}{(X_{02}-X_{01})X_{12}} + \frac{1}{(X_{03}-X_{01})X_{13}} \right) \\ + ({\rm cyc}\,123).
\end{multline}
By contrast, the surface recursion at 3-points gives (if we suppress the tadpole propagators $x_{ii}$)
\begin{align}
G &= \int_0^1 dt\, x_{01} G_{\rm tree}(01231) + ({\rm perm}\,123) \notag \\
&= \frac{1}{3}x_{01}x_{02}x_{03} + \frac{1}{2} x_{01}x_{02}x_{12} + \frac{1}{2} x_{03}x_{01} x_{31} + ({\rm cyc}\,123) \label{eq:notBCFW31} \\
& = x_{01}x_{02}x_{03} + \left( x_{01}x_{02}x_{12} + ({\rm cyc}\,123) \right). \notag
\end{align}
Equations \eqref{eq:BCFW31} and \eqref{eq:notBCFW31} after applying partial fractions to \eqref{eq:BCFW31}  in order to remove the spurious poles. But note that the surface recursion does not produce spurious poles as an intermediate step.

\paragraph{Shift sets.} Finally, we note that surface recursion allows us a much larger flexibility in choosing nice shift sets. One shift that works well for surface recursion is to shift the $n$ variables $X_{i,i+2}$ corresponding to the small cords of the disk. Every term in $G(1\cdots n; 0)$ contains at least one of these variables. However, if we shifted these variables by the BCFW shift, the resulting function $G(z)/z$ has a nontrivial contribution at $z=\infty$ from all the diagrams that only contain one $X_{i,i+2}$ propagator. This means that this is not a useful shift for the BCFW-like recursion.

\subsection{Matrix Model Correlators}\label{sec:matrix}
As a final application of surface recursion, we briefly consider how surface functions are related to matrix model correlators. Let $G_S$ be the surface functions for a polynomial interaction Lagrangian
\beq
{\cal L}_{\rm int} = \sum_{m=3}^\infty \frac{g_m}{m} \Tr (\phi^m).
\eeq
These can be computed from surface recursion with the boundary conditions
\beq\label{eq:boundarymm}
\left. G(12\cdots m) \right|_{x\rightarrow 0} = g_m
\eeq
for the $m$-point tree-level disks. But consider replacing all the variables $x_i$ in $G_S(x_1,x_2,\ldots)$ with a single variable, $x$. Then we obtain polynomials $G_S(x)$ in $x$, for every surface. By the chain rule, these polynomials satisfy a cut equation
\beq\label{eq:dertmm}
\partial_x G_S  = \sum_{i=1}^k \,G^\text{cut $i$}_S,
\eeq
where we sum over \emph{all} possible cuts. Surface recursion can be used to solve \eqref{eq:dertmm}, with boundary conditions \eqref{eq:boundarymm}, to obtain $G_S(x)$ for all surfaces. For example, for the 4-point tree-level disk,
\beq
G_{\text{4-pt}} = g_4 + 2 (g_3)^2 x,
\eeq
and, for the annulus with one marked point on each boundary,
\beq
G_\text{annulus} = \int_0^1 dt\, x( g_4 + 2 (g_3)^2 xt )= g_4 x + (g_3)^2 x^2.
\eeq
These polynomials are, in fact, contributions to the genus expansion of correlation functions for a Gaussian $U(N)$ matrix model whose partition function is
\beq
Z = \int \frac{ dM}{{\rm Vol}\,U(N)} \, \exp \left( - \frac{1}{2x} \Tr M^2 + \sum_{m=3}^\infty \frac{g_m}{m} \Tr M^m \right).
\eeq
We do not explore this connection further here, but note that the cut equation \eqref{eq:dertmm} appears to be very distinct from the loop equations/Virasoro constraints that are often used when solving for the genus expansion of matrix model partition functions.

\section{Uncolored Scalars}\label{sec:uncolored}
As explained in Section \ref{sec:surface}, the amplitudes of theories with an uncolored scalar $\phi$ also have a natural interpretation in terms of surfaces. The difference is that a $\phi$ propagator should be considered dual to a \emph{closed curve} that forms a closed loop on the surface. Whereas colored $\sigma$ propagators are labelled by curves that begin and end on boundaries and punctures. Despite this, we can still define surface functions $G_S$ for theories with an uncolored scalar in exactly the same way as before.

Introduce a variable, $z_\Delta$, for every $\MCG$ coset of closed curves $\Delta$ on a genus $g$ surface, $S$, with some $n$ marked punctures. This surface corresponds to a $g$-loop contribution to the amplitude for $n$ uncolored external states. For concreteness, consider just the cubic uncolored theory
\beq
{\cal L}_{\rm int} = \frac{1}{3!} \sigma^3,
\eeq
with no colored scalar fields. For this theory, the surface functions are
\beq\label{eq:GdefPsi}
G_S = \sum_G \prod_{\Delta \in G} z_\Delta,
\eeq
where we sum over all cubic (non-fat) graphs $G$. In the surface picture, this is a sum over all distinct (up to $\MCG$) decompositions of $S$ into 3-punctured spheres (this is also known as a \emph{pair of pants decomposition}). Remarkably, these surface functions also satisfy a cut equation
\beq\label{eq:cutPsi}
\partial_{z_\Delta} G_S = G_{S\,\text{cut along $\Delta$}}.
\eeq
This has exactly the same form as the cut equation studied earlier, in Section \ref{sec:cut}, for colored scalar theories. This is despite the fact that, on the surface, the two cut equations look very different: cutting along a closed curve produces new holes/punctures, whereas cutting along a non-closed curve produces new boundary edges.

As for colored theories, we can solve \eqref{eq:cutPsi} using surface recursion. Moreover, by choosing the boundary terms that appear in surface recursion, we can find surface functions for any uncolored scalar theory with arbitrary interaction Lagrangian. We illustrate this below at tree level, first for purely uncolored theories, and then for theories with coupled colored and uncolored scalars. At the end of the section we comment on the surface functions that appear at higher genus.

\subsection{Tree level amplitudes}
To compute uncolored scalar amplitudes at tree level, take $S$ to be the sphere with $n$ labelled punctures and write $G(12\cdots n)$ for the associate surface function. Each closed curve on $S$ is specified (up to $\MCG$) by a bipartition of $\{1,2,...,n\}$ into two disjoint subsets $A, A^c$ with $|A|, |A^c| \geq 2$. For a subset $A \subset \{1,\ldots,n\}$, let $z_A = z_{A^c}$ be the variable associated to the partition $A \cup A^c$. We recover the amplitude from $G(12\cdots n)$ by replacing $z_A$ with its associated propagator
\beq
z_A \rightarrow \frac{1}{\left(\sum_{i\in A} p_i \right)^2 + m^2}.
\eeq
If ${\cal L}_{\rm int}$ has $m$-point $\sigma$ interactions ${\cal L}^{(0,m)}$, surface recursion computes the amplitudes for the theory with boundary conditions
\beq
\left. G(12\cdots m) \right|_{z\rightarrow 0} = {\cal L}^{(0,m)}(12\cdots m)
\eeq
for the $m$-point surface functions. For example, $G(123) = {\cal L}^{(0,3)}(123)$ and, at 4-points, we find
\begin{multline}
G(1234) = z_{12} G(12p_{12}) G(p_{12}34) + z_{23} G(23p_{23})G(14p_{23}) \\ + z_{13} G(13p_{13})G(24p_{13}) + {\cal L}^{(0,4)}(1234),
\end{multline}
where $p_{ij}$ is the puncture produced by pinching $z_{ij}$, which has momentum $\pm (p_i^\mu+p_j^\mu)$. 

Finally, to illustrate how surface recursion can be useful even at tree level, consider again the special case of the cubic theory, ${\cal L}_{\rm int} = \sigma^3/3!$. In this case all the boundary terms ${\cal L}^{(0,m)}$ are $0$ for $m > 3$. Moreover, from the definition \eqref{eq:GdefPsi}, we see that every term in $G_S$ contains at least one variable of the form $z_{ij}$, corresponding to a curve that surrounds just \emph{two} marked punctures. So the amplitudes for the cubic theory can be computed using surface recursion by shifting just the $n(n-1)/2$ variables $z_{ij}$:
\begin{equation}
    G(12...n) = \sum_{ij} \int_0^1 dt  \, z_{ij} G(p, 1,2,\ldots\hat{i}\ldots \hat{j} \ldots n)(t) G(p,i,1j),
\end{equation}
where $p$ marks the new point created by cutting the curve $\Delta_{ij}$. Starting with the 3-point vertex, $G(123) = 1$, this gives
\begin{align}
    G(1234) = z_{12}+z_{23}+z_{13}
\end{align}
and, at 5-points,
\beq
    G(12345) =  \int_0^1 dt\, z_{12} G(0345)(t) + \cdots = \frac{1}{2} z_{12}(z_{34}+z_{35}+z_{45}) + \cdots,
\eeq
where we sum over all $10$ subsets $ij$ of $12345$. It is easy to check that $G(12345)$ is a sum over 15 terms, each corresponding to one of the 15 5-point Feynman graphs. This recursion is interesting because, at $n$-points, $G(12\cdots n)$ is a function of $2^{n-1} - (n+1)$ variables $z_A$, growing exponentially in $n$. Whereas, the recursion has $n(n-1)/2$ terms, which grows only as $\sim n^2$. See Appendix \ref{app:tree} for a comparison of tree-level surface recursion with Berends-Giele like recursions, for both uncolored and colored scalar theories.

\subsection{Coupling to a colored scalar}
The tree amplitudes for a theory of a colored scalar $\phi$ coupled to an uncolored scalar $\sigma$ can be computed using surface recursion, by computing surface functions $G_S$ that depend on both the $z_\Delta$ variables, associated to closed curves on $S$, and the $x_C$ variables, associated to non-closed curves. Now $G_S$ satisfies both the cut equation \eqref{eq:cutPsi} in the $z$ variables \emph{and} the cut equation in the $x$ variables.

Write $G_{n,m} = G(1\cdots n; 1, \cdots, m)$ for the surface function with $n$ external $\phi$ states (in a single trace ordering) and $m$ external (unordered) $\sigma$ states. The associated surface is an $n$-point disk with $m$ labelled punctures. This depends as before on the variables $z_A$, for each subset $A \subset \{1,\cdots,m\}$. Moreover, it depends on $x$ variables $x_{ij;A} = x_{ji; A^c}$, corresponding to the curves that connect the $\phi$-points $i$ and $j$, and partition the $\sigma$-punctures as $A\cup A^c$. After substituting the associated propagators, the functions $G_{n,m}$ are then precisely the tree amplitudes of the theory.

For an arbitrary Lagrangian, ${\cal L}_{\rm int}$, the interaction terms ${\cal L}^{(a,b)}$ (see Section \ref{sec:surface}) appear as boundary terms in the surface recursion. However, to illustrate the key idea, consider simply the cubic theory
\beq\label{eq:sigmaLint}
\Lint = \frac{1}{3} \Tr \phi^3 + \frac{1}{3!} \sigma^3 + \sigma \, \Tr \phi^2.
\eeq
The 3-point vertices give 
\beq
G(123)=1,\qquad G(\emptyset; 123) = 1,\qquad G(1;23)=0, \qquad G(12;3)=1,
\eeq
and then surface recursion (with no boundary terms) compute the higher-point cases. Explicitly, the cut equations that we are solving now read
\begin{align}
\partial_{x_{ij;A}} G_{n,m} & = G(i \cdots j; A ) G(j\cdots i; A^c),\\
\partial_{z_A} G_{n,m} & = G(1\cdots n; p A^c) \, G(\emptyset; pA).
\end{align}
At 4-points, we integrate to find, e.g.,
\beq
G(123;4) = x_{12;4} + x_{23;4} + x_{31;4},\qquad G(12;34) = x_{12;4} + z_{34},\qquad G(1;234) = 0,
\eeq
as well as the all-$\phi$ and all-$\sigma$ surface functions computed earlier
\begin{align}
G(1234) = x_{13} + x_{24},\qquad  G(\emptyset; 1234) = z_{12} + z_{13}+ z_{23}.
\end{align}
At 5-points, consider, for example, 
\beq
G(12;345) = \int dt\, x_{12;3} G(12;45) + x_{12;34} G(12;34) + z_{34} G(12;05) + \cdots.
\eeq
This gives
\beq
G(12;345) = x_{12;3}x_{12;34} + x_{12;3}z_{45} + x_{12;34}z_{34}+ z_{34} z_{345} + ({\rm perms}).
\eeq
To carry on the computation at higher points, in an efficient way, it is useful to choose a good shift. For example, we can shift just variables $x_{ii+2;\emptyset}$, $x_{ii+1;j}$ and $z_{jk}$, since every tree graph contributing to $G_{n,m}$ contains at least one of these variables. The variables in this set grow as $\sim n$ in $n$ and $\sim m^2$ in $m$, whereas the number of terms in $G_{n,m}$ grows exponentially in both $n$ and $m$.

\subsection{Higher genus and symmetry factors}\label{sec:Psiloop}
At higher genus, $G_S$ does not compute loop integrands for theories with uncolored scalars. This is because it is not possible to assign propagators to the $x$ and $z$-variables: curves that are $\MCG$-equivalent, and so correspond to a single $x$ or $z$ variable, correspond to propagators that carry different momenta. However, analogous to Section \ref{sec:matrix}, these functions can be used to find matrix model correlation functions — now for a model with a $U(N)$ matrix $M$ coupled to a scalar.

We briefly outline the computation of these higher genus surface functions and, for simplicity, focus just on the cubic scalar theory ${\cal L}_{\rm int} = \sigma^3/3!$ for a single uncolored scalar field. Let $G_g(1\cdots n)$ be the surface function for the genus $g$ surface with $n$ marked punctures. For example, $G_1(1)$ is the function for the torus with one marked puncture, labelled $1$. This surface has only one $\MCG$ class of closed curves, which are the genus-reducing curves that cut the surface to a 3-punctured sphere. Call the variable associated to these curves $z$. Then
\beq
\partial_z G_1(1) = G(001) = 1,
\eeq
where we label the two new punctures created by cutting $z$ with $0$. So
\beq
G_1(1) = z
\eeq
and this term corresponds to the single tadpole diagram for the $\sigma^3$ theory. 

Now consider $G_1(12)$, for the torus with two punctures. In addition to the curves, $z$, that cut the torus to a sphere, we also have one new variable $z_{12}$ associated to the curves that separate the two punctures from the rest of the torus. The cut equations are then
\beq\label{eq:genusPsiex}
\partial_z G_1(12) = G(0120),\qquad \partial_{z_{12}} G_1(12) = G(012) G_{1}(0). 
\eeq
Earlier we computed that, $G(0120) = z_{01} + z_{02} + z_{12}$. However, the curves $\Delta_{01}$ and $\Delta_{02}$ on the punctured sphere correspond to the genus-reducing curves $\Delta$ on the torus. So we write
\beq
G(0120) = 2z + z_{12}.
\eeq
The surface recursion then gives
\beq
G_{1}(12) = \frac{1}{2} z( 2z+z_{12}) + \frac{1}{2} z_{12} z = z^2 + z z_{12},
\eeq
and these two terms correspond to the two 2-point 1-loop diagrams (the bubble and the tadpole) of the $\sigma^3$ theory. 

The above recursion reproduces our original definition of the uncolored surface functions $G_S$ for $\sigma^3$ theory, \eqref{eq:GdefPsi}, which does \emph{not} include symmetry factors. We mention in passing that we can define surface functions $\widetilde{G}_S$ for the cubic theory that include symmetry factors. To do this we modify the definition of surface functions for the case when the punctures have repeated (indistinguishable) labels. The separating variables are still defined by subsets $A$ of the set of labels. But when there are repeated labels, there are fewer than $n(n-1)/2$ distinct subsets $A$. For example, if we have four punctures labelled $\{0,0,1,2\}$, the variables $z_{01}$ and $z_{02}$ correspond to identical partitions. This motivates us to consider replacing $G(0120) = 2 z_{01} + z_{12}$ with
\beq
\widetilde{G}(0120) = z_{01} + z_{12}.
\eeq
Then solving the cut equations, \eqref{eq:genusPsiex}, but now using $\widetilde{G}(0120)$, gives
\beq
\widetilde{G}_1(12) = \frac{1}{2} z^2 + z z_{12}.
\eeq
Here, the term $z^2$ corresponding to the bubble diagram is multiplied by the bubble's symmetry factor, $1/2$. Whereas the tadpole has symmetry factor $1$, which is the coefficient of $zz_{12}$.

\section{Discussion and Outlook}

In this paper we introduced the natural notion of stringy surface functions, and we explored their properties in the leading ``low-energy" limit, where they can be determined recursively using the cut equation. 
We stress that the cut equation gives a type of recursion that is very different from the recursions we are accustomed to in field theory. 
One natural comparison is with recursion relations that arise from the use of Cauchy's residue theorem in conjunction with the factorization and vanishing at infinity properties of integrands --- such as the BCFW \cite{Britto_2005W} recursion for Yang-Mills and gravity, or triangulations of polytopes for scalar theories \cite{2017ABHY,halo,simpleforms}. 
These recursion always introduce spurious poles, that only cancel in the full sum. 
By contrast we never see spurious poles when solving the cut equation.  

Another comparison is with Berends-Giele recursion \cite{BERENDS1988759}, which gives a simple way of partitioning tree-level diagrams into sets, counting each of them once. 
Even in the simplest cases, Berends-Giele considers two (or more) cuts at each step, while the cut equation only ever sees single cuts. Moreover, solving the cut equations does not simply partition all diagrams into groups, but rather overcounts them in systematic fashion that involves \emph{fewer} terms at each step in the recursion. 
This overcounting is reflected in the fact that while we never see spurious poles, we do see ``spurious fractions": seemingly wrong rational coefficients, that add up to the correct coefficients when all terms are summed.

In this paper we have motivated and explained surface functions in a self-contained way, without referring to the machinery of the Feynman fan, $u$-variables and the curve integral formalism \cite{counting1,counting2}. 
But surface functions and the cut equation were forced on us in the course of studying the curve integrals and a tropical version of the Mirzakhani integration scheme \cite{mirzakhani2007}, that has been discussed extensively in \cite{counting1,counting2}.
In fact, remarkably these ideas lead to a well-defined family of (non-unique) integrands for general scalar theories \emph{beyond} the planar limit, even when including uncolored scalar fields. These non-unique integrands can also be computed recursively using ideas very similar to the cut equation in this paper. There is also no obstruction to extending the recursion in this paper to theories with multiple species of particles, and to theories with spinning particles. We will report on these applications elsewhere. 

There are many avenues for direct extensions of this paper.  Among other things, the cut equations can be efficiently solved to count diagrams (with or without symmetry factors) at all orders in the topological expansion. It is interesting to understand these classic combinatorial problems from this new point of view.  
As we have emphasized, the main purpose in life of the cut equation is to transparently account for the combinatorics of Feynman diagrams, viewed as polyangulations of surfaces.   In  applications to field theory, the cut equation is not sensitive to the ``on-shell" nature of amplitudes; indeed the cut equation can be used to sum diagrams for off-shell correlators coming from general Lagrangians as well. This is seen vividly in the way we handle general momentum-dependent interactions --- the variables ``upstairs", in the numerator, are kept distinct from those of the poles, and we only set them equal to each other at the end of the calculation. However, ``on-shell" objects enjoy many additional wonderful properties. For instance, they are field redefinition invariant, and in the case of gauge theories they enjoy on-shell gauge invariance,  related to the stunning simplification of amplitudes relative to the naive expectation from Feynman diagrams. It is therefore natural to ask whether and how the extra simplifications of on-shell physics can be captured by differential equations. Is there a simple way to see why field redefinitions leave amplitudes invariant? Are there differential equations for the surface functions of interesting theories, like the NLSM and Yang-Mills, when the kinematic numerators and denominators are identified from the outset? 

Relatedly, recent work has shown that the description of scattering amplitudes in terms of curves on surfaces allows us to define and calculate ``canonical'' loop integrands that manifest important properties of the amplitude, such as the Adler zero for the NLSM. These integrands can be naturally extracted from deformations of the $\mathrm{Tr}(\phi^3)$ integrand --- which itself can been calculated by means of surface recursion. It remains to be seen whether these ``perfect'' integrands can also be efficiently calculated using the cut equation by choosing suitable boundary conditions. We will leave this question to future work.

Finally, the way in which the \emph{stringy surface functions} generalize matrix model correlators is particularly fascinating. Matrix models have been studied for decades in contexts ranging from statistical mechanics to the earliest avatars of holography in string theory and quantum gravity. So it is a pressing question to understand the physical interpretation of the stringy surface functions, which are a double-generalization of matrix models to finite values of $\alpha^\prime$ and general kinematic $X_C$ variables.  In this paper we have turned off $\alpha^\prime$, but considered the case of general $X_C$. A first obvious step to understanding these objects is to think about the opposite extreme, keeping all $X_C$'s equal but turning on $\alpha^\prime$, defining a ``stringy" extension matrix models. The $\alpha'$ expansions of these new functions have interesting connections to number theory. 
In particular, they connect Weil-Petersson volumes \cite{mirzakhani2007} to the rich structure of multiple zeta values appearing in perturbative string amplitudes \cite{brown2010periods}.

This deserves to be studied in much greater depth. Is there a physical interpretation of these functions? And what is the finite $\alpha^\prime$ generalization of the cut equation?

\paragraph{Acknowledgements.}
The work of N.A.H., H.F. and G.S. is supported by the DOE (Grant No. DE-SC0009988), further support was made possible by the Carl B. Feinberg cross-disciplinary program in innovation at the IAS. N.A.H. and H.F. are also supported by the  European Union (ERC, UNIVERSE PLUS, 101118787).  N.A.H. is further supported by the Simons Collaboration on Celestial Holography. HF is also supported by the Sivian Fund. 
The work of G.S. is part of the PositiveWorld project funded by the European Union’s Horizon 2023
research and innovation programme under the Marie Skłodowska-Curie grant agreement 101151760.
Views and opinions of the authors expressed are those of the author(s) only and do not necessarily reflect those of the European Union or the European Research Council Executive Agency. Neither the European Union nor the granting authority can be held responsible for them.

\appendix

\section{Surface Functions with $\alpha^\prime$ corrections}\label{app:ap}
In this section we describe how the surface functions are related to integrals on moduli spaces. For each surface, $S$, we can naively consider a string-theory-like integral
\beq\label{A:naive}
\alpha^{\prime \, {\rm dim}\, {\cal M}(S)} \int \frac{\omega}{\MCG} \prod_C (u_C)^{\alpha^\prime \tilde{X}_C},
\eeq
which is an integral over the moduli space ${\cal M}(S)$, or, equivalently, over the Teichmuller space ${\cal T}(S)$ modulo the $\MCG$. Here, $\omega$ is a volume form on ${\cal T}(S)$, and the integrand is given as a product over all curves $C$ (up to homotopy) on the surface $S$. For each homotopy class of curves, $u_C$ is a function on ${\cal T}(S)$ that roughly corresponds to a cross-ratio of lengths on the surface (see \cite{counting1}).

The problem with \eqref{A:naive} is that the integrand is not $\MCG$ invariant, and so quotienting by $\MCG$ does not make sense. This is because, in  \eqref{A:naive}, there is one variable $\tilde{X}_C$ for each \emph{homotopy class} of curves, but non-homotopic curves can still be equivalent under the $\MCG$ action. 

To solve this, consider how the $\MCG$ partitions the set of all curves into cosets. Picking some coset representative $C$ for each coset, we can replace the $\tilde{X}_C$ variables with a single variable, $X_C$, for each coset. This gives a well-defined integral on the moduli space,
\beq
{\cal G}_S(\alpha^\prime, X) = \alpha^{\prime \, {\rm dim}\, {\cal M}(S)} \int \frac{\omega}{\MCG} \prod_C \left( \prod_{\gamma\in\MCG} u_{\gamma C} \right)^{\alpha^\prime \,X_C},
\eeq
where the product in the integrand is now over $\MCG$ cosets of curves (and for convenience we identity each coset with some coset representative $C$). The resulting string-like function, ${\cal G}_S(\alpha^\prime, X)$, is a function of the variables $X_C$, one for each $\MCG$ coset. The surface functions for $\Tr\phi^3$ theory are the $\alpha^\prime \rightarrow 0$ limit of these ${\cal G}_S(\alpha^\prime, X)$,
\beq
\lim_{\alpha^\prime \rightarrow 0} {\cal G}_S(\alpha^\prime, X)  = G_S(X).
\eeq
We do not show this in detail here, but note that it is useful to use the properties of the \emph{tropicalization} of the $u_C$ (see again \cite{counting1}).

Taking a derivative of  ${\cal G}_S(\alpha^\prime, X)$ with respect to one of the variables, $X_D$, gives
\beq
\partial_{X_D} {\cal G}_S = \alpha^{\prime \, {\rm dim}\, {\cal M}(S)} \int \frac{\omega}{\MCG} \left( \sum_{\gamma \in \MCG} \alpha^\prime\, \log u_{\gamma D} \right) \prod_C \left( \prod_{\gamma\in\MCG} u_{\gamma C} \right)^{\alpha^\prime \,X_C}. 
\eeq
But, by an orbit-stabilizer argument,
\beq\label{A:penult}
\partial_{X_D} {\cal G}_S = \alpha^{\prime \, {\rm dim}\, {\cal M}(S)} \int \frac{\omega}{{\rm Stab}\, D} \left( \alpha^\prime\, \log u_{ D} \right) \prod_C \left( \prod_{\gamma\in\MCG} u_{\gamma C} \right)^{\alpha^\prime \,X_C},
\eeq
where ${\rm Stab}(D)$ is the stabilizer subgroup of $\MCG$ that leaves the curve $D$ fixed. In fact, ${\rm Stab}(D)$ can be identified by the $\MCG$ of the surface obtained by cutting $S$ along $D$. Using this observation, we claim that it can be shown that, in the $\alpha^\prime\rightarrow 0$ limit, the RHS of this equation becomes
\beq
RHS \rightarrow - \frac{1}{X_D^2}\, G_{S\,\text{cut along}\, D}.
\eeq
So, taking the $\alpha^\prime\rightarrow 0$ of \eqref{A:penult} gives
\beq
\partial_{X_D} G_S = - \frac{1}{X_D^2}\, G_{S\,\text{cut along}\, D},
\eeq
which is the cut equation after substituting $X = 1/x$.

\section{Tree level surface recursion and Berends-Giele}\label{app:tree}
We give here some further details on how surface recursion can be used to compute tree amplitudes in theories of colored and/or uncolored scalars, and compare this to the Berends-Giele-like recursions that can also be used to compute these tree amplitudes. \cite{BERENDS1988759,mafraBG}

\subsection{Colored scalar theories}
Suppose we wish to compute the tree-level amplitudes $G_n = G(12\cdots n)$ for a theory of a single colored scalar $\phi$. If all interaction vertices ${\cal L}^{(k)}$ vanish for $k > m$, for some $m$, then we can make an intelligent choice of shift set to compute the $n$-point amplitudes. For $n \leq m$, we include \emph{all} variables in the shift set. However, for $n>m$ it is only necessary to shift a subset of the variables. In particular, take the set of $x_{ij}$ for $i,j$ not more than $m-1$ steps apart on the disk: i.e. satisfying $|j-i| \leq m-1$, where $|j-i| = {\rm min}(j-i, n-j+i)$ is the difference between $i$ and $j$, cyclic mod $n$. Every term in $G_n$ contains at least one variable from this set. So, for $n> m$, surface recursion using this shift computes $G_n$ as
\beq\label{B:Gnsmall}
G_n = \int_0^1dt\, \sum_{|j-i|\leq m-1} \,x_{ij} \, G_n^{{\rm cut}\,ij} (t).
\eeq
For sufficiently large $n$, there are $n(m-2)$ summands in this sum. (Note that if $n=3$, this shift set is comprised of the variables $x_{i,i+2}$.)

It is interesting to compare \eqref{B:Gnsmall} to the Berends-Giele-like recursion for the tree amplitudes of the same class of theories. Assuming that the interactions do not vanish, ${\cal L}^{(k)} \neq 0$, for all $k\leq m$, a Berends-Giele recursion for the $n$-point amplitude of this theory is a sum over
\beq
f(n,m) = \sum_{k=3}^{m} {n-2 \choose k-2} 
\eeq
summands --- since for each $k$ between $3$ and $m$, we must sum over the number of ways to distribute $n-1$ ordered external points among $k-1$ lower-point amplitudes. $f(n,m)$ has power-law like growth in $n$ for large $n$, and is dominated by $n^{m-2}$. At one extreme, if $m=3$, the surface recursion has $n$ summands, whereas Berends-Giele recursion has $n-2$, so the number of summands is similar in both cases. But if $m > 3$, the Berends-Giele recursion is a sum over $\sim n^{m-2}$ summands, whereas the surface recursion, \eqref{B:Gnsmall}, is a sum over $n(m-2)$ summands, which grows much more slowly in $n$.

Finally, at the other extreme, consider the case that the interaction vertices ${\cal L}^{(m)}$ are $\neq 0$ for \emph{all} $m$. Then the Berends-Giele recursion is a sum over
\beq
 \sum_{k=3}^{n} {n-2 \choose k-2} = 2^{n-2} -1
\eeq
summands, which grows exponentially in $n$. Whereas, the surface recursion at $n$-points is a sum over all chords of the disk, giving $n(n-3)/2$ summands, which grows quadratically in $n$.

\subsection{Uncolored scalar theories}
Consider now the tree-level amplitudes $G_n = G(1,2,\cdots,n)$ for some theory of a single uncolored scalar $\sigma$. For any $\Lint$, the tree amplitudes can be computed via surface recursion using an appropriate shift set. For each nonzero interaction term ${\cal L}^{(k)}$ in the Lagrangian, we include in our shift set the ${n \choose k-1}$ variables $z_A$ corresponding to curves that separate $k-1$ of the punctures from the others. For large $n$, ${n \choose k-1}$ grows as $n^{k-1}$. So the total number of variables in the shift set is polynomial in $n$ and dominated by $n^{m+1}$, where $m$ is the valence of the highest-valence nonzero interaction in $\Lint$. 

Berends-Giele-like recursions are not often applied to uncolored theories. But we can still apply it to compute the amplitudes $G_n$. For each $k$ with ${\cal L}^{(k)}\neq 0$, we sum over all ways to distribute $n-1$ \emph{unordered} points amoung $k-1$ lower-point amplitudes (themselves unordered): so a $k$-valent vertex contributes 
\beq
\sum_{a_i \geq 0}\frac{1}{k!} {n-1-k \choose a_1, \cdots, a_{k}}
\eeq
summands to this Berends-Giele-like recursion. The number grows exponentially as $\sim k^n$ with $n$. Whereas the number of terms in the surface recursion for $G_n$ grows polynomially in $n$.

\section{Implementing surface recursion for planar integrands}\label{app:code}
A Mathematica notebook for computing surface functions using surface recursion is included with this manuscript. The focus of the notebook is to compute $n$-point $\ell$-loop surface functions, $G_{n,\ell} = G(1\cdots n; p_1;\cdots; p_\ell)$, for an arbitrary theory of a colored scalar $\phi$. These $G_{n,\ell}$ reproduce the planar integrands of the theory after substituting the variables with kinematic data (as explained in the main text). In this section we collect some observations about surface recursion that proved useful for implementing the recursion for these planar integrands. 

\paragraph{A simple planar shift set.} As discussed in Appendix \ref{app:tree}, there is a natural shift set to use to compute the tree-level functions, $G_{n,0}$, for any Lagrangian, $\Lint$, which includes chords of the n-point disk up to some length. Moreover, if $\Lint$ has only finitely many terms, this shift set grows only linearly in $n$ for large $n$ (whereas the number of terms in $G_{n,0}$ grows exponentially in $n$). 

To compute $G_{n,\ell}$ for $\ell \geq 1$, a particularly convenient shift is comprised of all the variables $x_{ip}$ that connect an external point $i$ on the boundary and a puncture $p$. Every term in $G_{n,\ell}$ must include at least one of these variables, so this is a good shift for surface recursion:
\beq\label{B:rec}
G_{n,\ell} = \int_0^1 dt\, \sum_{i,p} x_{ip}\, G^{\,\text{cut along}\, x_{ip}}_{n,\ell},
\eeq
where we sum over all $n\ell$ variables $x_{ip}$. Moreover, the cuts along these variables give $\ell-1$ loop surface functions with $n+2$ marked points:
\beq
\partial_{x_{ip}} G_{n,\ell} = G(1\ldots i, p, i \ldots n; \,p_1;\ldots; \hat{p};\ldots; p_\ell).
\eeq
This is particularly nice. It means that the functions $G_{n,\ell}$ can be computed entirely recursively from other planar surface functions $G_{n',\ell'}$ with $\ell' < \ell$ and $n' + 2 \ell' = n + 2\ell$. Ultimately, $G_{n,\ell}$ can be computed using this recursion in terms of tree-level functions $G_{n',0}$ with $n' = n + 2\ell$.

As an aside, a simple generalization of this shift is useful beyond the planar case. For a genus $g$ surface with (possibly multiple) boundaries, fix some boundary and consider all curves with an end-point on that boundary which are non-separating (i.e. do not cut the surface into two surfaces). This set of curves can be used as a shift set to compute the non-planar surface functions for the theory. 

\paragraph{Simpler variables.} In the main text, we defined $G_{n,\ell}$ to be a function of all $\MCG$ classes of curves. This is not always the most convenient way to define $G_{n,\ell}$ in practice. Consider the curves connecting a pair $i,j$ of points on the boundary. They are not all $\MCG$ equivalent. These curves cut the punctured disk into two parts, and partition the punctures in two. Two curves that partition the punctures in different ways are $\MCG$ equivalent. So we are lead to introduce variables $x_{ij}[A]$ for every possible subset $A \subset \{p_1, \ldots, p_L\}$ of the punctures. 

The problem is that, when we come to compute the integrand from $G_{n,\ell}$, all of these variables $x_{ij}[A]$, for different subsets $A$, are assigned to the \emph{same} propagator factor, since they have the same momentum (e.g. given by $P = z_i^\mu - z_j^\mu$ in dual momentum variables). So it appears we are doing a lot of ``excess book-keeping'' to compute the planar integrands.

This can be resolved by identifying all of these variables and replacing them with a single variable $x_{ij}$. This does not change anything about the recursion, \eqref{B:rec}, defined above. The advantage of making this identification is that $G_{n,\ell}$ then depends on the same number of variables as there are distinct propagator factors in the final integrand.

However, it is worth noting that, after making this identification, the planar surface functions satisfy a modified cut equation for cuts along the $x_{ij}$:
\beq\label{B:nopart}
\partial_{x_{ij}} G_{n,\ell} = \sum_A G(i\cdots j; A) G(j\cdots i; A^c),
\eeq
where the sum is over all possible subsets $A \subset \{ 1,\ldots, \ell\}$ of the punctures.  For example, we can check that the 2-loop tadpole surface function (after these identifications),
\beq
G(1;2;3) = x_{12}x_{13}^2x_{23} + x_{11} x_{12} x_{13}^2  + x_{11}^2 x_{12}x_{13} + x_{22}x_{12}x_{23} + (2\leftrightarrow 3),
\eeq
satisfies \eqref{B:nopart} for the derivative with respect to $x_{11}$:
\beq
\partial_{x_{11}} G(1;2;3) = G(1;2) G(11;3) + G(1;3) G(11;2)
\eeq
where
\beq
G(1;2) = x_{12},\qquad G(11;3) = x_{13}^2 + 2 x_{11}x_{13}.
\eeq

\paragraph{Vector version of the cut equation.} To implement the surface recursion on a computer, it is not convenient to view each step in surface recursion as an integral. It is better to express the recursion as a vectorial statement. For example, consider the case of shifting \emph{all} $x$ variables at each step in the recursion. Rather that writing $G(t)$ as a polynomial in $t$, it can be convenient for computer calculations to write it is a vector, $G^a$, so that $G(t) = \sum_a G^a t^a$. Then the cut equation,
\beq
\partial_t G(t) = \sum_{a=1} a t^{a-1} G^a = \sum_{\text{cut $x$}} x \sum_{a=0} t^a G^a_{\rm cut},
\eeq
can be written as (for $a \geq 1$)
\beq
G^{a} = \sum_{x}  \frac{x}{a} \, G_{\text{cut $x$}}^{a-1},
\eeq
which is a vectorial version of the recursion. The $G^0$ entry of the vector is the boundary term for the recursion, and is theory-dependent.

\bibliographystyle{unsrt}
\bibliography{refs}

\end{document}